# Errors and their mitigation at the Kirchhoff-law-Johnson-noise secure key exchange

Yessica Saez, Laszlo B. Kish

Department of Electrical and Computer Engineering, Texas A&M University, College Station, TX 77843-3128, USA

## Abstract

A method to quantify the error probability at the Kirchhoff-law-Johnson-noise (KLJN) secure key exchange is introduced. The types of errors due to statistical inaccuracies in noise voltage measurements are classified and the error probability is calculated. The most interesting finding is that the error probability decays exponentially with the duration of the time window of single bit exchange. The results indicate that it is feasible to have so small error probabilities of the exchanged bits that error correction algorithms are not required. The results are demonstrated with practical considerations.

## Introduction

### 1.1 The KLJN secure key exchange

In today's era, network security has become one of the most important aspects in everyday life. Whether it is a large, small, private, or a government organization, it is very important to focus on security, especially when the data being sent, received, or stored contain confidential, sensitive information, such as personal information.

In private-key based secure communication, the two communicating parties (Alice and Bob) generate and share a secure key, which is typically represented by a random bit sequence. It is important to note that the security of a communication cannot be better than the security of the exchange of the key it uses. During this key exchange, the eavesdropper (often referred to as Eve) is continuously monitoring the related data. In today's Internet-based secure communications, typically a software–based key generation and distribution is utilized. However, in this method the whole information about the secure key is publicly available [1] and Eve's access to this information is limited only by her computational power. In other words, this method provides only a (*computationally*) *conditional* security level, which represents a non-future-proof-security [2-4]. It means that with a sufficiently enhanced computation power or an efficient future algorithm, Eve may be able to crack the key and all the information in the communication may become accessible.

Therefore, scientists and researchers have been working on exploring proper laws of physics to find new key exchange schemes where the information that can be measured by Eve is zero. Particularly, they have been exploring key exchange schemes where the amount of information extracted by Eve does not depend on her computational power. When the security measures are determined at Eve's maximal ability (limited only by the laws of physics and the protocols working conditions), that is referred as *unconditional security*, a term that is often interchanged with *information theoretic security* [1]. Information theoretic (unconditional) security can be *perfect* if Eve can extract no information, or *imperfect*, if Eve can extract only a small, commonly accepted amount of information. (This is allowed for practical purpose because this small information leak can further be decreased by privacy amplification, if the fidelity of the key exchange between Alice and Bob is good enough.) These terms are often misunderstood, and it is a frequent mistake in claims to misuse *unconditional security and imply perfect security* by that.

It is important to emphasize that the goal to generate/distribute a perfectly secure key is similar to approaching infinity. Perfectly secure key distribution of a key of finite length can never be reached with a real physical system within a finite duration of time. However, it is one of the goals of physical informatics to find out schemes that can arbitrarily approach (though never reach) perfect security [2].

The earliest and most famous scheme based on the laws of physics that is claiming unconditional security is the Quantum Key Distribution (QKD) [5]. The information theoretic security of this scheme is usually based on the assumption that Eve's actions will disturb the system (in accordance with the theory of quantum measurements and the no-cloning theorem) and cause errors, which uncover the eavesdropping. Note, there are some promising non-QKD initiatives that involve new types of quantum effects [6, 7].

At the fundamental side, there are ongoing debates between experts about the reachable levels of security in QKD [8-12]. At the practical side, there are some issues associated with this scheme, such as range, price, and robustness. Moreover, it is interesting to note that recently all the commercial QKD devices and many laboratory devices have been cracked by quantum-hacking [13-27]. While most of these practical weaknesses seem to be design flaws, not fundamental security problems; they still mean that current practical QKD has yet conditional security: the conditions are that Eve is not knowledgeable enough or she does not have the proper hardware to utilize the design flaws for an attack. The impressive list of papers [13-27] shows that there are enough knowledgeable Eves out with sufficient resources at the moment.

Until 2005 QKD was the only accepted scheme that was able to offer a key exchange with information theoretic security in the ideal (mathematical) situation. In 2005, the Kirchhoff-Law-Johnson-(like)-Noise (KLJN) secure key distribution was introduced [28], where the term "totally secure" was used instead of the correct "perfectly secure" expression. Later (2006), the KLJN system had been built and demonstrated [29]. KLJN is also a key exchange scheme with information theoretic security [3] and it is based on Kirchhoff's Loop Law of quasi-static electrodynamics and the Fluctuation and Dissipation theorem of statistical physics. Its security against passive attacks is ultimately based on the Second Law of Thermodynamics [28], which means that it is as hard to crack the key exchange as to build a perpetual motion machine (of the second kind). At practical conditions it uses enhanced (electronically generated) Johnson noise with high noise temperature, where quasi-static and thermodynamic aspects must be emulated as exactly as possible in order to approach perfect security.

First, we present a brief description (based on [2-4, 28]) of the working principle of the KLJN system. The core KLJN system, without the defense circuitry against invasive attacks and vulnerabilities represented by non-ideal building elements is shown in the following figure.

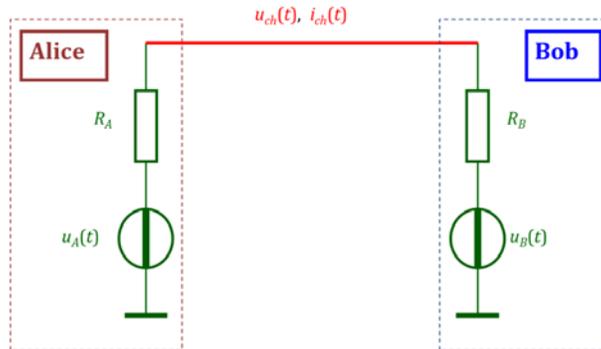

**Figure 1. Outline of the core KLJN secure exchange scheme [2-4, 28] without the defense elements against active (invasive) attacks or attacks utilizing non-ideal components and conditions.**

The core KLJN channel, see Fig. 1, is a wire line to which Alice and Bob connect randomly selected resistors $R_A$ and $R_B$, respectively, where $R_A, R_B \in \{R_0, R_1\}$. $R_0$ represents the low (0) bit and $R_1$ the high (1) bit, respectively [28]. At the beginning of each bit exchange period, BEP, (also called KLJN clock period), Alice and Bob, who possess identical pairs of the resistors $R_0$ and $R_1$, randomly select and connect one of these resistors. The Gaussian voltage noise generators represent either the Johnson noises of the resistors or external noise generators delivering band-limited white noise with publicly known bandwidth and effective noise temperature $T_{eff}$ [2, 3, 28, 29]. The noise voltages of Alice and Bob are $u_A(t)$ and $u_B(t)$, respectively, where $u_A(t) \in \{u_{0,A}(t), u_{1,A}(t)\}$ and $u_B(t) \in \{u_{0,B}(t), u_{1,B}(t)\}$ yield a channel noise voltage $u_{ch}(t)$ between the wire line and the ground and a channel noise current $i_{ch}(t)$ in the wire.

Alice and Bob measure the mean-square noise voltage and/or current amplitudes, that is $\langle u_{ch}^2(t) \rangle$ and/or $\langle i_{ch}^2(t) \rangle$, within the BEP in the line. Thus, by applying Johnson's noise formula and Kirchhoff's loop law the theoretical prediction is that the mean-square noise voltage and current (i.e. the integral of the corresponding power spectral densities [2,28]) for a given channel noise bandwidth $B_{KLJN}$ and temperature $T_{eff}$ are given as follows:

$$\langle u_{ch}^2(t) \rangle = S_{u,ch}(f) B_{KLJN} = 4kT_{eff} R_{\parallel} B_{KLJN}$$
$$\langle i_{ch}^2(t) \rangle = S_{i,ch}(f) B_{KLJN} = 4kT_{eff} \frac{1}{R_{loop}} B_{KLJN} \quad , \tag{1}$$

where $\langle \; \rangle$ represents ideal (infinite-time) time average, $S_{u,ch}(f)$ is the power density spectrum of channel voltage noise, $S_{i,ch}(f)$ is the power density spectrum of channel current noise, $k$ is the Boltzmann constant, $R_{\parallel} = R_A R_B / (R_A + R_B)$ and $R_{loop} = R_A + R_B$.

Ideally, by comparing the result of the accurate measurement of the mean-square channel voltage or current with the corresponding theoretical value in Eqs. 1, the total loop resistance will be publicly known. Alice and Bob know their own resistor values and thus they can deduce that resistance value from the loop resistance to learn the resistance at the other end. Consequently, they can distill the actual bit value at the other side of the wire.

If Alice and Bob use the same resistance values, Eve can also recognize that bit situation because the total resistance is either the lowest or the highest value of the three possible resistance values. Thus, the resistor situations $(R_0, R_0)$ and $(R_1, R_1)$ represent a non-secure bit exchange since Eve can also find out the resistors values, their exact locations, and the status of the bits. On the other hand, the cases $(R_0, R_1)$ and $(R_1, R_0)$, which yield identical mean-square noise in the line, represent a secure bit exchange situation because Eve is unable to locate the resistors, therefore, she cannot decide if Alice (and Bob) has a bit 1 or 0. This security is provided by the Second Law of Thermodynamics, which prohibits any directional information concerning the resistors at the two sides in thermal equilibrium [2,28]. In other words, it is as difficult to extract these secure bits by Eve as to build a perpetual motion machine (of the second kind). In conclusion, on average, 50% of the bits can be kept because they are secure. The other 50% of the bits representing the non-secure situations is discarded by the protocol.

Note: the securely exchanged bits have opposite values at Alice and Bob, thus they must publicly agree which one of them will invert the exchanged bit to have identical keys at the two ends.

The fully armed KLJN system is secure even against the man-in-the-middle-attack [30]. One of the important potential applications [32] is to integrate the KLJN system on computer chips and provide unconditional security within computers and high-security instrumentations where the processors, hard drives, keyboards, etc. would secure their communications by keys shared via the KLJN protocol. Another, potential application is, at a much greater scale, to build a network of KLJN systems utilizing already existing wire lines [4, 33, 34], particularly, realizing and unconditionally secure "smart grid" [4] (advanced electrical power distribution network).

**1.2 Known attack types**

Below, based on [2], we briefly survey all the published attack types. Due to the simplicity of the KLJN system, there are very few attack types available. The method of comparing the instantaneous values of voltage and current at the two ends and discarding risky 01/10 bits [28, 30, 31] (not discussed here in details) protects against all these types of attacks. But even without discarding the risky bits, passive attacks by Eve utilizing non-idealities suffer from weak signal-to-noise ratio due to poor statistics, see below.

A practically unimportant but theoretically valid type of attack was shown by Hao [36] who pointed out that the non-ideal situation of different temperatures could separate the noise levels of the 01 and 10 bit situations, thus they could give out some information to Eve. In a response by Kish [37], it was pointed out that practical problems of accuracy do not challenge the conceptual security of ideal schemes and was estimated that, even at practical situations, the information leak is negligible due to this attack. Later, it was shown in the experimental paper of Mingesz et al. [29] that a modest 14-bit accuracy of temperatures (noise generators) practically prohibit Eve to extract any useful information (with information leak less than $10^{-10}$) by utilizing the Hao attack.

Scheuer and Yariv [38] analyzed the case of non-zero wire resistance where the mean-square voltages are different at the two ends in the case of the 01 and 10 bit situations. However, their calculation was incorrect including the physical units of some of the main results. Kish and Scheuer [39] carried out new, correct calculations and showed that the actual effect is about 1000 times weaker than predicted by Scheuer and Yariv. Earlier, Kish pointed out [37] in his response to [36] that at similar conditions Eve's statistic was very poor and the extracted information was practically miniscule even without the defense of discarding the risky bits. This claim was experimentally verified by Mingesz et al. [29], who showed that at clock period of 50 times of the noise correlation time, $R_0 = 2000 \, \Omega$, $R_1 = 9000 \, \Omega$, and wire resistance $200 \, \Omega$, the information leak of exchanged raw bits to Eve was 0.19% while the fidelity between Alice and Bob was 99.98%. These results indicate that the key exchange has excellent fidelity even without error correction and that the security can be made reasonably good even without dropping the risky 01/10 bits (after current/voltage comparison at the two ends) and without privacy amplification [29].

Liu [41] used a cable simulator to evaluate the impact of delays and reflections on the security. He obtained the surprising results that, with the experimental parameters [6], Eve successfully guessed 70-80% of the key bits. In a critical study of Lui's simulations, Kish and Horvath [31] pointed out that the chosen wave impedances of the simulated cable to reach these results were unphysical: for example, a center wire diameter of 1 millimeter implies a coaxial cable with outer diameter of 28000 times greater than the size of the known universe.

Observing transients after switching the resistors has been mentioned as a potential source of information leak; however, so far they have never been utilized. During the experimental studies, the noise was ramped up at the beginning of the clock period and ramped down at the end, thus the switching of resistors took place when the voltage and currents were zero in the line.

Note, a fully transient-free protocol is described in a recent work [48].

According to [40], one of the most efficient attack types would be utilizing capacitive currents via the cable capacitance, though it has never been tested. Mingesz et al. [29] showed a hardware based defense "capacitance killer" against this attack. Ultimately, the method of discarding the risky bits after current/voltage comparison at the two ends [28, 30, 31] and/or, in the case of negligible error probability, privacy amplification [35] are the tools to approach perfect security.

### 1.3 Bit errors in the KLJN key exchange

Due to the finite duration $\tau$ of the bit exchange period BEP, the measurement results of mean-square amplitudes have statistical inaccuracies. The duration $\tau$ of the BEP must be long-enough compared to the correlation time of the noise (approximately the reciprocal noise-bandwidth $B_{KLJN}^{-1}$) to achieve a satisfactory statistics and safely distinguish between the different resistor situations. Still, with a low probability, these uncertainties can trigger a bit error.

In the experimental demonstration Mingesz et al. [29] were able to optimize the system to have a fidelity of 99.98% (error probability 0.02%) however no mathematical analysis or design tools have been shown to address this problem. Therefore, our goal in this paper is to classify the different types of bit errors in the ideal KLJN system and analyze their impact.

## Discussions and Results

### 2.1 KLJN Errors

In this "startup" paper about error analysis, we assume the ideal situation of the KLJN system where all the non-ideal features of real systems are neglected. The error analysis of non-ideal systems will be done in future works.

Bit errors occur when the actual value of the mean-square noise results in an incorrect bit interpretation. Figure 2 represents the mean-square channel noise voltage levels, where $\langle\ \rangle_\tau$ indicates finite ($\tau$) time average implying random fluctuations (statistical errors) around the real mean-square value.

The 11 bit situation (when Bob's and Alice's chosen resistors are $R_1$ and their noise voltages are $u_{1,A}(t)$ and $u_{1,B}(t)$, respectively) results in the mean-square channel noise voltage $\langle u_{11}^2(t)\rangle_\tau$. Similarly the 01/10 situations yield $\langle u_{01/10}^2(t)\rangle_\tau$ and the 00 bit arrangement results in $\langle u_{00}^2(t)\rangle_\tau$. The threshold values $\Delta_1$ and $\Delta_2$ provide the boundaries to interpret the measured mean-square channel voltage over the $\tau$ time window, see Fig. 2. The bit interpretation is 00 when $\langle u_{ch}^2(t)\rangle_\tau < \langle u_{00}^2(t)\rangle + \Delta_1$, and 11 when $\langle u_{ch}^2(t)\rangle_\tau > \langle u_{11}^2(t)\rangle - \Delta_2$. The secure bit situation 01/10 is interpreted when $\langle u_{00}^2(t)\rangle + \Delta_1 \le \langle u_{ch}^2(t)\rangle_\tau \le \langle u_{11}^2(t)\rangle - \Delta_2$.

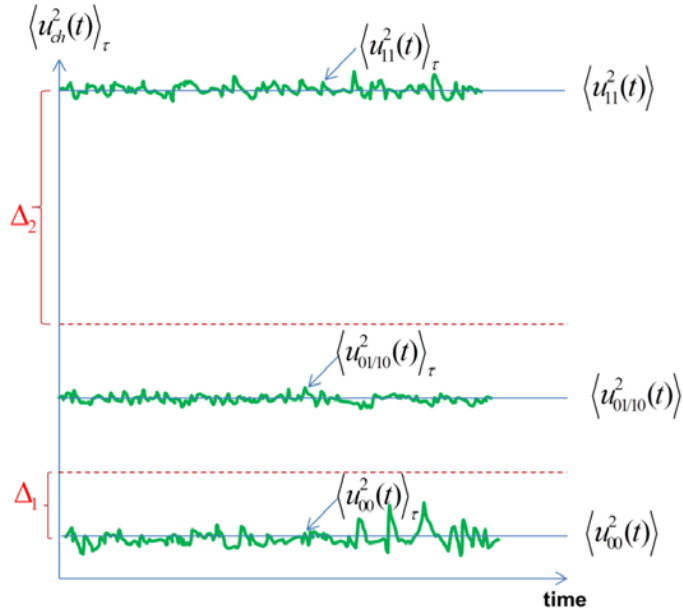

**Figure 2. Illustration of the fluctuations of the finite-time mean-square voltage levels around their exact value and thresholds for interpretation (the scale is arbitrary).**

An example for a bit error is the rare occurrence when the finite-time mean-square voltage of the 00 case, $\langle u_{00}^2(t)\rangle_\tau \geq \langle u_{00}^2(t)\rangle + \Delta_1$, is interpreted as the 01/10 bit situation, which is incorrect and an example of a bit error.

The different types of errors are shown in Table 1.

**Table 1. Types of errors in the KLJN bit exchange.**

|  |  | Actual Situation | | |
|---|---|---|---|---|
|  |  | 00 | 11 | 01/10 |
| **Measurement Interpretation (Decision)** | 00 | Correct (no error) | Error, Removed (automatically) | Error, Removed (automatically) |
|  | 11 | Error, Removed (automatically) | Correct (no error) | Error, Removed (automatically) |
|  | 01/10 | Error * (probability?) | Error * (probability?) | Correct (no error) |

**\*The rest of the paper addresses these errors and their probability.**

Some of the errors situations, as shown in Table 1, are considered to be self-corrected by the protocol. This is because, as aforementioned, the 00 and 11 bit situations are discarded.

The rest of the paper is dealing with the analysis of errors indicated with * in Table 1.

**2.2 Error probabilities in the KLJN scheme**

Alice and Bob can calculate the total resistance in the system by measuring the mean-square noise voltage and/or current amplitudes, that is, $\langle u_{ch}^2(t)\rangle_\tau$ and/or $\langle i_{ch}^2(t)\rangle_\tau$. Below we evaluate the errors in the former case while the case of current-based evaluation can be done in a very similar fashion.

## 2.2.1 Error probability due to inaccuracies in noise voltage measurements

### a) Probability of the 00 ==> 01/10 type errors

Let $R_0 = R$ and $R_1 = \alpha R$ with $\alpha \gg 1$. Note, the choice of $\alpha$ does not influence the resulting equations but it determines the upper limit at choosing the values of $\Delta_1$ and $\Delta_2$ (see Eqs. 1). Then, the mean-square channel noise voltage for infinite-time average at the 00 bit situation is given as:

$$\langle u_{00}^2(t) \rangle = S_{00}(f) B_{KLJN}, \tag{2}$$

where $S_{00}(f) = S_{u,ch}(f)$ at the bit situation 00. Because $R_\parallel = R/2$, from Eqs. 1, we obtain:

$$\langle u_{00}^2(t) \rangle = 2kT_{eff} R B_{KLJN} \tag{3}$$

During the BEP, only the duration $\tau$ is available for Alice, Bob and Eve to determine the mean-square channel noise because, after that, a new bit exchange begins. The block diagram of the measurement process is shown in Fig. 3.

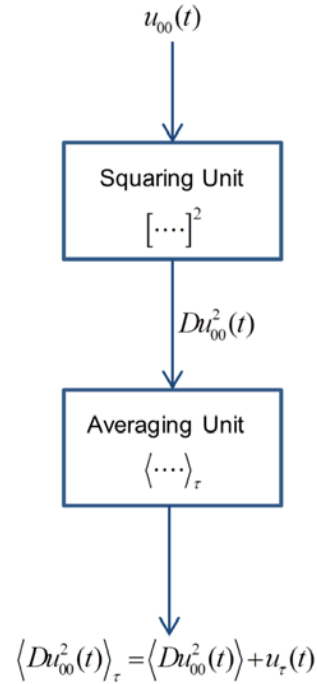

**Figure 3. Illustration of the measurement process at 00.**

The channel voltage enters into a squaring unit. At its output, the signal is still voltage (because it is a voltage-signal-based electronics) and the numerical value of its instantaneous amplitude is equal to the square of the instantaneous amplitude of the input voltage. This fact is mathematically expressed by $Du_{00}^2(t)$, where $D = \frac{1}{\text{Volt}}$ is the transfer coefficient of the device to provide a Volt unit also for the square [42]. After averaging for the finite-time $\tau$ duration, the obtained measurement result is $\langle Du_{00}^2(t)\rangle_\tau = \langle Du_{00}^2(t)\rangle + u_\tau(t)$, where the averaging can be represented by a low-pass filtering with cut-off frequency $f_B \approx 1/\tau$.

While $u_{00}^2(t)$ is not Gaussian, its finite-time average $u_\tau(t)$ is Gaussian with high accuracy due to the Central Limit Theorem, because $\tau$ is much longer than the correlation time of the AC component $u_{2,00}(t) = Du_{00}^2(t) - \langle Du_{00}^2(t)\rangle$ of $Du_{00}^2(t)$, as $f_B \ll B_{KLJN}$. The probability of 00 ==> 01/10 type errors is the probability that the AC component *remaining* after the finite-time average of $Du_{00}^2(t)$ defined as $u_\tau(t) = \langle Du_{00}^2(t)\rangle_\tau - \langle Du_{00}^2(t)\rangle$ is beyond the threshold: $u_\tau(t) > \Delta_1$. This can be evaluated by the error function, however, requires numerical integration.

To have an analytic formula, which is a good approximation and has the exact scaling in the small error probability limit, that is, when $u_\tau(t) \ll \Delta_1$ is satisfied, we can use Rice's formula [43, 44] of threshold crossing frequency, see similar solutions for estimating the probability of thermal noise induced switching errors [45-47]. The estimation of error probability is based on the fact that, in the small error limit, the probability of repeated threshold crossings within the correlation time of the band-limited noise converges to zero. The correlation time of $u_\tau(t)$ is also equal to $\tau$ thus each threshold crossing (in a chosen but fixed direction) indicates an independent error. The ratio of the mean threshold crossing frequency $\nu(\Delta_1)$ and $\tau$ is a good estimation of the error probability in this limit [45, 46]. We compared the predictions of the Rice formula with the prediction based on numerically evaluated error function and found that the Rice formula gave always more pessimistic error estimation. The variation of the threshold resulted in changing the error probability prediction by the Rice formula and the error function by factors of $\sim 10^{43}$ and $\sim 10^{44}$, respectively. In the large error probability situation, the Rice formula predicted about 2 times greater error while, in the low error probability situation, about 18 times greater error. This is a negligible difference not only due to the $10^{43}$ - $10^{44}$ variation during the study but also because the exact error probability slightly depends on the fine details of the protocol not discussed here. To have analytic error estimation, we proceed as follows.

According to Rice, the mean frequency $\nu$ of crossing the level $\Delta_1$ by a Gaussian with power density spectrum $S_\tau(f)$ is given as:

$$\nu(\Delta_1) = \frac{2}{\hat{u}_\tau} \exp\left(\frac{-\Delta_1^2}{2\hat{u}_\tau^2}\right) \sqrt{\int_0^\infty f^2 S_\tau(f) df} \tag{4}$$

where $S_\tau(f)$ is the power density spectrum of $u_\tau(t)$ and $\hat{u}_\tau$ is its RMS value, $\hat{u}_\tau = \sqrt{\int_0^\infty S_\tau(f) df}$.

For normalization purposes, we choose the $\Delta_1$ threshold level as a fraction of the *measured* mean-square channel noise, where the transfer coefficient $D$ of the squaring unit is also taken into the account:

$$\Delta_1 = \beta \langle D u_{00}^2(t) \rangle = \beta D S_{00}(f) B_{KLJN} , \quad \text{where } 0 < \beta < 1 . \tag{5}$$

According to [42], the power density spectrum, $S_{2,00}(f)$, of the AC component $u_{2,00}(t)$ of the (non-averaged) $u_{00}^2(t)$ is given as (note typos of missing factor of 2 in Eqs. 6 and 7 in [42], see Fig. 4):

$$S_{2,00}(f) = 2D^2 B_{KLJN} S_{00}^2(f)(1 - \frac{f}{2B_{KLJN}}) \quad \text{for} \quad 0 \le f \le 2B_{KLJN} \quad \text{and} \quad S_{2,00}(f) = 0 \quad \text{otherwise} \tag{6}$$

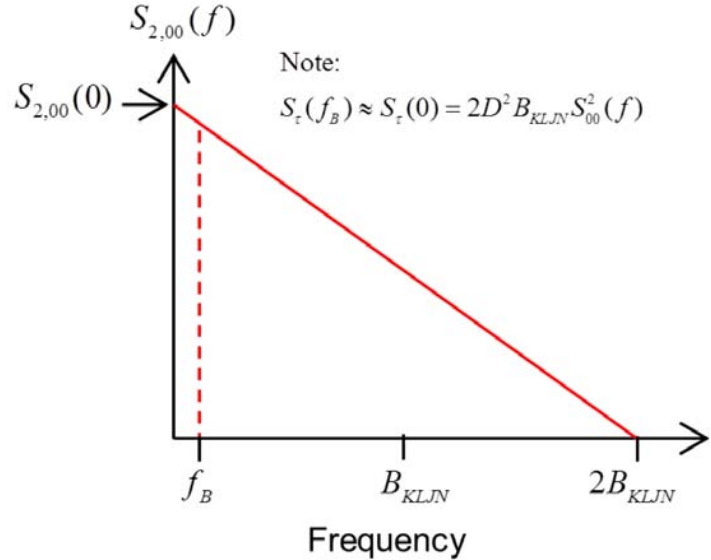

**Figure 4. Power Spectral Density (PSD) of the product of two independent noises.**

The low-pass filtering effect of the time averaging cuts off this spectrum for $f > f_B$ but keeps the $S_{2,00}(f)$ spectrum for $f < f_B$. Because $f_B \ll B_{KLJN}$, the value of $S_{2,00}(f)$ within the $f_B$ frequency band can be approximated by its maximum, $S_\tau(f) \approx S_{2,00}(0)$. Figure 5 summarizes these findings.

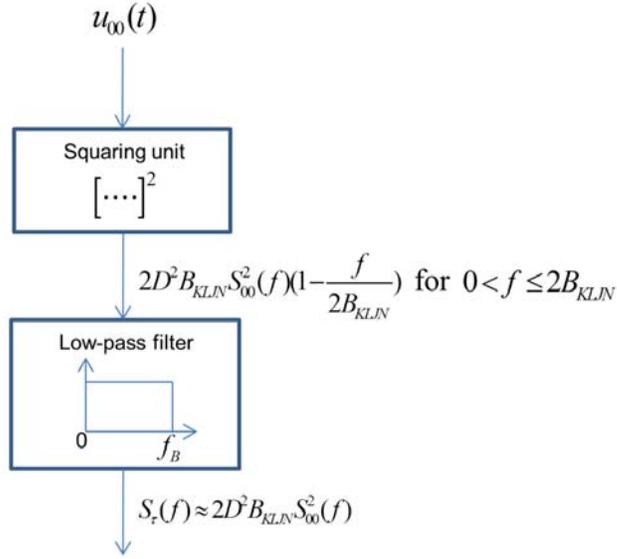

**Figure 5. Spectra at the 00 bit situation.**

Let us suppose that $B_{KLJN} / f_B = \gamma$. Then

$$\hat{u}_\tau = \sqrt{\int_0^\infty S_\tau(f)df} \approx \sqrt{f_B S_{2,00}(0)} = \sqrt{2D^2 \gamma f_B^2 S_{00}^2(f)} \quad , \tag{7}$$

see text above and Figure 3 for explanation of the approximation. The frequency $v_\uparrow(\Delta_1)$ of unidirectional level crossings is half of the level crossing frequency predicted by the Rice formula:

$$v_\uparrow(\Delta_1) = \frac{1}{\hat{u}_\tau} \exp\left(\frac{-\Delta_1^2}{2\hat{u}_\tau^2}\right) \sqrt{\int_0^\infty f^2 S_\tau(f) df} \quad , \tag{8}$$

where

$$\Delta_1 = \beta D S_{00}(f) \gamma f_B \tag{9}$$

From Eqs. 7 and 9, we obtain

$$v_\uparrow(\Delta_1) = \frac{f_B}{\sqrt{3}} \exp\left(\frac{-\beta^2 D^2 S_{00}^2(f) \gamma^2 f_B^2}{4 D^2 \gamma f_B^2 S_{00}^2(f)}\right) = \frac{f_B}{\sqrt{3}} \exp\left(\frac{-\beta^2 \gamma}{4}\right) \tag{10}$$

In the high threshold situation the errors follow a Poisson statistics, thus the error probability during a time interval is equal to the expected numbers of errors within this interval provided this number is much less than 1.

Thus the probability $\varepsilon_{00}$ of 00==>01/10 type of errors in the case of $\varepsilon_{00} \ll 1$ is:

$$\varepsilon_{00} \approx \nu_\uparrow(\Delta_1)\tau \approx \frac{\nu_\uparrow(\Delta_1)}{f_B} = \frac{1}{\sqrt{3}}\exp\left(\frac{-\beta^2\gamma}{4}\right) \tag{11}$$

It is important to realize that the error probability is an exponential function of the parameters. The $\gamma$ parameter (which is proportional to the length of time average) is particularly important because it is not limited in size.

**b) Probability of the 11 ==> 01/10 type errors**

We can follow the same procedure as above. Instead of $\beta$ we introduce $\delta$ with similar meaning, see Fig. 2 and Eq. 5:

$$\Delta_2 = \delta\langle Du_{11}^2(t)\rangle = \delta DS_{11}(f)B_{KLJN} = \delta\gamma DS_{11}(f)f_B, \qquad 0 < \delta < 1 \tag{12}$$

where $\Delta_2$ is the threshold for the 11==>01/10 type errors and $S_{11}(f)$ is the channel noise spectrum at the 11 bit situation.

The same type of calculations as given above yields the probability $\varepsilon_{11}$ of 11==>01/10 type errors:

$$\varepsilon_{11} = \frac{\nu(\Delta_2)}{f_B} = \frac{1}{\sqrt{3}}\exp\left(\frac{-\delta^2\gamma}{4}\right) \text{ for } 0 < \delta < 1 \tag{13}$$

The error probability is again an exponential function of the parameters.

**2.3 Illustration of the results with practical parameters**

To demonstrate the results, we assign possible practical values to the parameters.

For $\gamma = 100$ and $\beta = 0.5$ (a choice allowed due to the $\alpha \gg 1$ condition, see Eqs. 1) the bit error probability $\varepsilon_{00}$ is:

$$\varepsilon_{00} = \frac{1}{\sqrt{3}}\exp\left(\frac{-\beta^2\gamma}{4}\right) \approx 0.001, \tag{14}$$

which is a value near to the experimental value (0.0002) obtained in [29] with the same $\gamma = 100$ value (note the $\beta$ value is not available in [29] however the $\beta = 0.5$ choice is a practical one). If this value is too large, just by increasing the $\gamma$ parameter (and the time average window $\tau$) by a factor of 2, and in this way slowing down the bit exchange by the same factor, will result in the square of the above error probability value:

$$\varepsilon_{00} \approx 10^{-6} \quad , \tag{15}$$

which is satisfactory for most applications. It is important to note that no error correction algorithm is used for this error reduction.

## Methods and Conclusions

We have classified and analyzed the types of errors of bit exchange between Alice and Bob in the KLJN secure key exchange. Some types of errors are automatically removed by the original protocol. We mathematically analyzed the error probabilities and their dependence on the KLJN parameters of the errors that are not removed by the protocol. We identified the important parameters and the results show that the error probability decays exponentially by increasing these parameters. The most important of such parameters is the duration $\tau$ of key exchange because its value is not limited. The results indicate that it is reasonable to achieve error probabilities that are small enough to avoid the need for error correction algorithms.

Further open questions are how to combine current and voltage measurements to further reduce these errors and what is the error situation in the new advanced KLJN protocols proposed recently [48].

## Acknowledgements

Related discussions with Elias Gonzalez and Claes-Göran Granqvist are appreciated.